\documentclass[amsmath,10pt,twocolumn,pre,aps,superscriptaddress]{revtex4-2}
\usepackage[utf8]{inputenc}
\usepackage{hyperref}
\usepackage{graphicx}
\usepackage{ragged2e}
\usepackage{subcaption}
\usepackage{tabularx}
\usepackage{arydshln}
\usepackage{multirow}
\usepackage{booktabs}
\usepackage[percent]{overpic}
\usepackage{amsmath,bm, amssymb}
\usepackage{uri}
\usepackage[dvipsnames]{xcolor}

\begin{document}
\title{Odd Polycatenanes Tank-Tread in Strong Shear Flow}
\author{Ali Seyedi}
\affiliation{Department of Chemical Engineering and Materials Science, Wayne State University, 5050 Anthony Wayne Drive, Detroit, Michigan 48202, USA}

\author{Charles W. Manke}
\affiliation{Department of Chemical Engineering and Materials Science, Wayne State University, 5050 Anthony Wayne Drive, Detroit, Michigan 48202, USA}

\author{Alex Albaugh}
\email{aalbaugh@wayne.edu}
\affiliation{Department of Chemical Engineering and Materials Science, Wayne State University, 5050 Anthony Wayne Drive, Detroit, Michigan 48202, USA}
\affiliation{NSF-Simons National Institute for Theory and Mathematics in Biology, Chicago, Illinois 60611, USA}

\begin{abstract}
Polycatenanes are mechanically interlocked polymers composed of ring molecules linked through mechanical bonds. 
Here, we investigate the single-molecule dynamics of linear polycatenanes under strong steady shear flow using coarse-grained Brownian dynamics simulations with hydrodynamic interactions. 
We identify a stable tank-treading state in which individual rings rotate continuously while the overall polymer remains highly extended and aligned with the flow direction, adopting conformations typically associated with extensional flow.
Remarkably, tank-treading is observed almost exclusively in odd polycatenanes, polycatenanes with an odd number of rings. 
We attribute this odd-even effect to the orientation of the terminal rings within the flow-gradient plane, where they act as anchors that stabilize polymer extension, a configuration accessible only to odd polycatenanes. 
Furthermore, our simulations and a Markov state model demonstrate that this dynamic behavior remains stable over long timescales. 
Finally, we show that tank-treading polycatenanes can be used to fully extend other molecules in strong shear flow, hinting at applications of this behavior to manipulate molecules in rotational flows.
\end{abstract}

\maketitle

\section{Introduction}
\label{sec:intro}
\begin{figure}
    \centering

    \begin{subfigure}{0.4\textwidth}
        \centering
        \begin{overpic}[width=\textwidth]{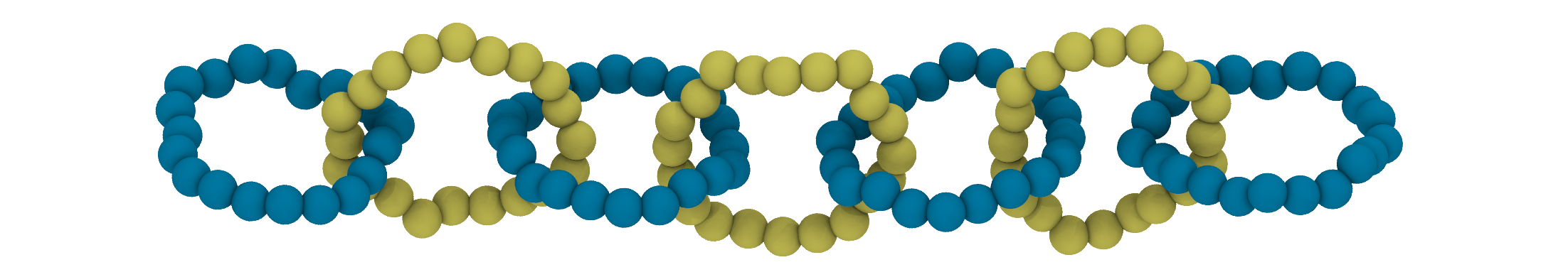}
            \put(2,18){\normalsize\sffamily (a)}
        \end{overpic}
        \label{fig:polycatenane}
    \end{subfigure}

    \vspace{0.5em}

    \begin{subfigure}{0.4\textwidth}
        \centering
        \begin{overpic}[width=\textwidth]{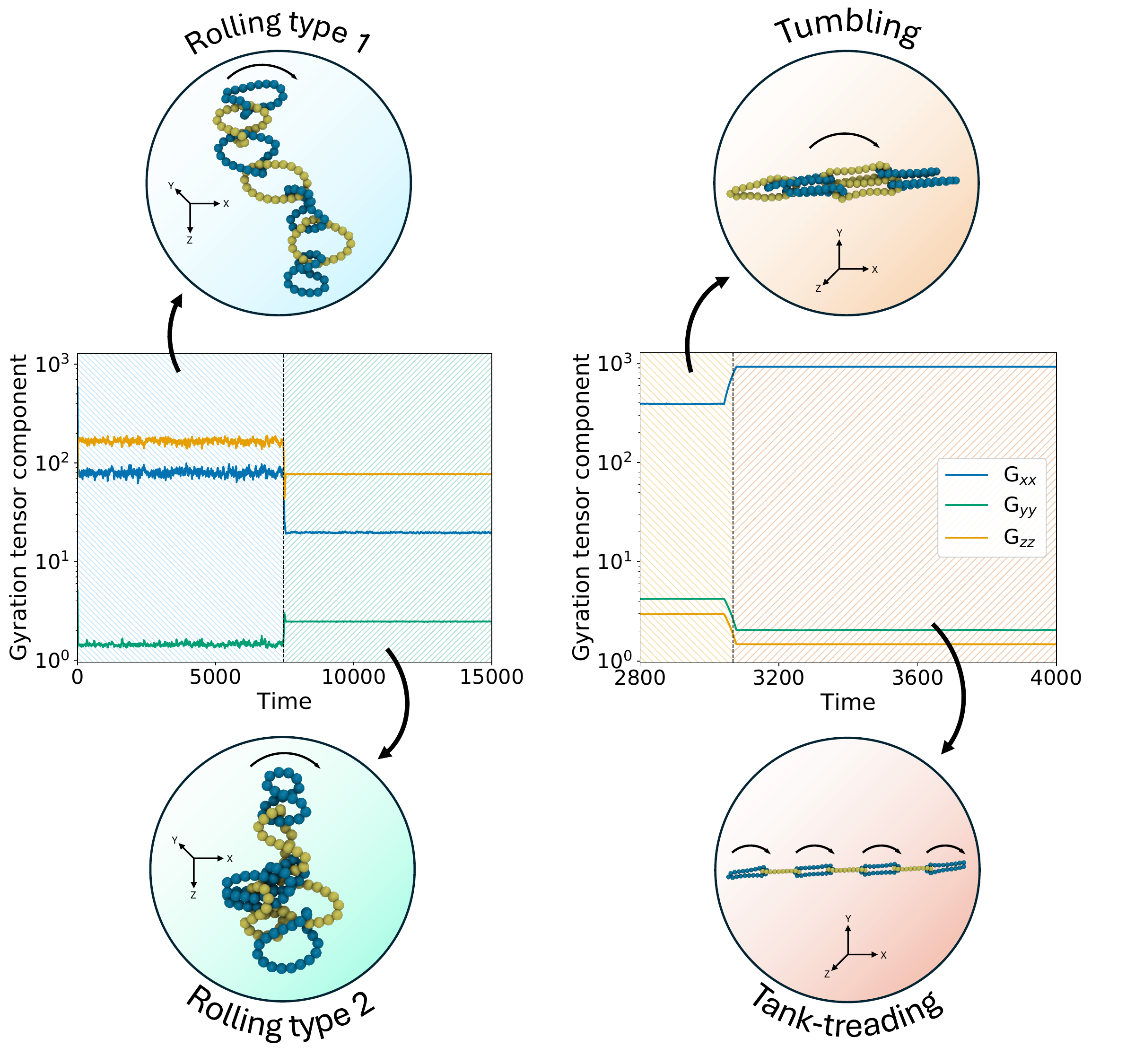}
            \put(2,92){\normalsize\sffamily (b)}
        \end{overpic}
        \label{fig:classes_bottom}
    \end{subfigure}
    \captionsetup{justification=justified,singlelinecheck=false}

    \caption{\justifying
  \small{(a) A polycatenane is a linear chain of interconnected rings.
  Here we illustrate a [7]catenane.
  In our model all beads are physically the same, but we color beads in alternating rings blue and yellow for visual clarity.
  (b) In strong shear flow, a polycatenane can exhibit several behaviors: rolling type 1 (blue background), rolling type 2 (green background), tumbling (orange background), and tank-treading (red background).
  We classify these behaviors based on the distinct values of the gyration tensor components (\(G_{xx}\), \(G_{yy}\), \(G_{zz}\)) in each mode.
  In rolling type 1, the catenane is extended and rolls in the vorticity (\(z\)) direction.
  Rolling type 2 is similar to rolling type 1, but the polymer is more compact.
  When tumbling, the polymer undergoes large-scale rotation in the flow-gradient (\(xy\)) plane.
  When tank-treading, the catenane adopts a steady extended configuration while alternating rings rotate in the flow-gradient plane.}}
  \label{fig:intro}
\end{figure}

Linear polycatenanes are chains of interlocked molecular rings~\cite{gil2015catenanes}, illustrated in Fig.~\ref{fig:intro}a.
From the earliest interlocking of two molecular rings into [2]catenanes in the 1960s~\cite{wasserman1960preparation,schill1964preparation}, synthetic techniques developed to interlock 3~\cite{schill1969catenanes}, 4~\cite{amabilino1994twostep}, 5~\cite{Amabilino1994Olympiadane}, 6 and 7~\cite{amabilino1998oligocatenanes} rings into linear chains.
Recently, supramolecular toruses have been assembled to form linear polycatenanes of up to 22 rings~\cite{datta2020self} and linear polycatenanes of up to 2,800 molecules rings have been reported~\cite{wang2025fully}.
These synthetic advancements demonstrate the wide range of possible polycatenane structures across length scales and numbers of rings.

The interlocked structures impart polycatenanes with unique properties.
Compared to typical linear polymers, polycatenanes have slower relaxation at short length scales~\cite{rauscher2018topological}, increased \(\theta\)-temperature~\cite{dehaghani2020effects}, increased tensile modulus~\cite{pakula1999simulation}, and faster relaxation in the melt state~\cite{rauscher2020dynamics}.
Compared to bonded ring polymers, they have greater elasticity~\cite{chen2023topological} and in moderately strong shear flow they exhibit fundamentally different tumbling dynamics~\cite{farimani2024effects}.
Under strong confinement, polycatenanes show unique overstretching compared to linear polymers~\cite{chiarantoni2023linear} and slower translocation through pores compared to ring polymers~\cite{caraglio2017driven}.
Our recent simulation study demonstrated that, when subject to moderate strength shear flow, polycatenanes exhibit suppressed shear thinning compared to linear polymers~\cite{seyedi2026mechanically}.
Overall, the topological links of polycatenanes lead to distinct properties and dynamics.

Here, using Brownian dynamics simulations, we demonstrate another unique behavior of polycatenanes.
In shear flows at a high shear rate, polycatenanes can exhibit a stable ``tank-treading'' behavior.
In this mode, the rings of the polycatenane align alternately between the flow-gradient and flow-vorticity planes. 
Rings in the flow-gradient plane rotate at a constant rate.
Overall, the polymer adopts a steady-state extended conformation in the flow direction.
We only observe this tank-treading behavior in odd polycatenanes, polycatenanes with an odd number of rings.

Tank-treading has previously been observed in ring polymers in shear flow~\cite{chen2013tumbling,chen2013effects,chen2015conformations,lang2014dynamics,liebetreu2020hydrodynamic}, blends of ring and linear polymers in pressure-driven Poiseuille flow~\cite{wang2024threading}, and branched ring polymers~\cite{wang2023single}.  
Additionally, tank-treading-like behavior has been observed in star polymers in shear flow, where the overall configuration is steady, but arms rotate about the center~\cite{ripoll2006star}.
It is also a well-known behavior in larger soft objects such as droplets and capsules~\cite{ripoll2006star}, as well as vesicles, and cells~\cite{abkarian2007swinging,tsubota2010effect,fischer1978red,keller1982motion}.

The presence of tank-treading in polycatenanes is notable because the polymer adopts a stable, extended configuration, which is not typical of many polymers in shear flow, which tend to tumble~\cite{de1974coil, smith1999single, teixeira2005shear}.
We further demonstrate that by attaching polycatenanes to either end of a typical linear polymer (which undergoes stretch-tumble cycles in shear flow), the linear polymer can also be stretched into a stable extended configuration.
This observation hints at possible applications of polycatenanes to stretch molecules into steady configurations in complex flows.

\section{Methods}
We simulated single polymers in the dilute solution limit using an established coarse-grained bead-spring model for mechanically interlocked polymers (MIPs) ~\cite{rauscher2020thermodynamics,becerra2024single}. 
Non-bonded interactions are described by the Weeks–Chandler–Andersen potential ~\cite{weeks1971role} and bonds by FENE springs ~\cite{kremer1990dynamics}. 
To avoid mechanical bond crossing at high shear, non-bonded interactions and bond strengths are scaled up by a factor of 100 and the FENE maximum extension is reduced by 20\% compared to typical Kremer-Grest parameterizations~\cite{kremer1990dynamics}, resulting in a stiff polymer.

The position of each bead (\(i=1,2,...N\)) evolves according to the following stochastic differential equation~\cite{ermak1978brownian,ottinger1996stochastic,rotne1969variational,yamakawa1970transport}:
\begin{equation}
d\mathbf{r}_{i}=\left[\mathbf{v}_{i}+\frac{1}{k_B T}\sum_{j=1}^{N}
\mathbf{D}_{ij}\cdot \mathbf{f}_{j}\right]dt+\sqrt{2}\sum_{j=1}^{N}
\mathbf{B}_{ij}\cdot d\mathbf{w}_{j}.
\end{equation}
In this overdamped Brownian dynamics approach, the position of a bead (\(\mathbf{r}_{i}=[x_i,y_i,z_i]^\top\)) evolves based on the solvent flow velocity (\(\mathbf{v}_{i}\)), the forces due to interactions between beads (\(\mathbf{f}_{i}\)), and a stochastic term capturing thermal fluctuations.
Here \(x\), \(y\), and \(z\) directions are flow, gradient, and vorticity directions, respectively.
In shear flow with shear rate \(\dot{\gamma}\), the flow velocity is given by \(\mathbf{v}_{i}=[\dot{\gamma} (y_i - y_{\mathrm{COM}}), 0, 0]^\top\), centered at the polymer center of mass \cite{miao2017iterative}. 
The stochastic contribution is drawn from a Wiener process \(d\mathbf{w}_{i}\), where \(\mathbf{w}_{i}\) is a vector of independent Wiener processes with increments having zero mean and variance $dt$.
The Rotne-Prager-Yamakawa mobility tensor \(\mathbf{D}_{ij}\) models hydrodynamic interactions between beads and \(\mathbf{B}_{ij}=\sqrt{\mathbf{D}_{ij}}\).

As illustrated in Fig.~\ref{fig:intro}a, we construct an [n]catenane from \(n\) interlocked rings, each formed with 20 particles connected with FENE bonds.
All simulations were carried out at a shear rate of $\dot{\gamma}=343$.
Uncertainties were obtained by parametric bootstrap~\cite{efron1992bootstrap}.
Non-dimensionalized variables were used in the simulations.
Additional details of the model, non-dimensionalization, and the computational approach are described in the SM.
%, except for cases without hydrodynamic interactions where a shear rate of $\dot{\gamma}=100$ was used to ensure numerical stability.
%Polymer stress is computed via the Kramers expression~\cite{bird1977dynamics,Graham_2018,bosko2008effect}. 

\section{Results}

\begin{figure}
    \centering
    \includegraphics[width=0.49\textwidth]{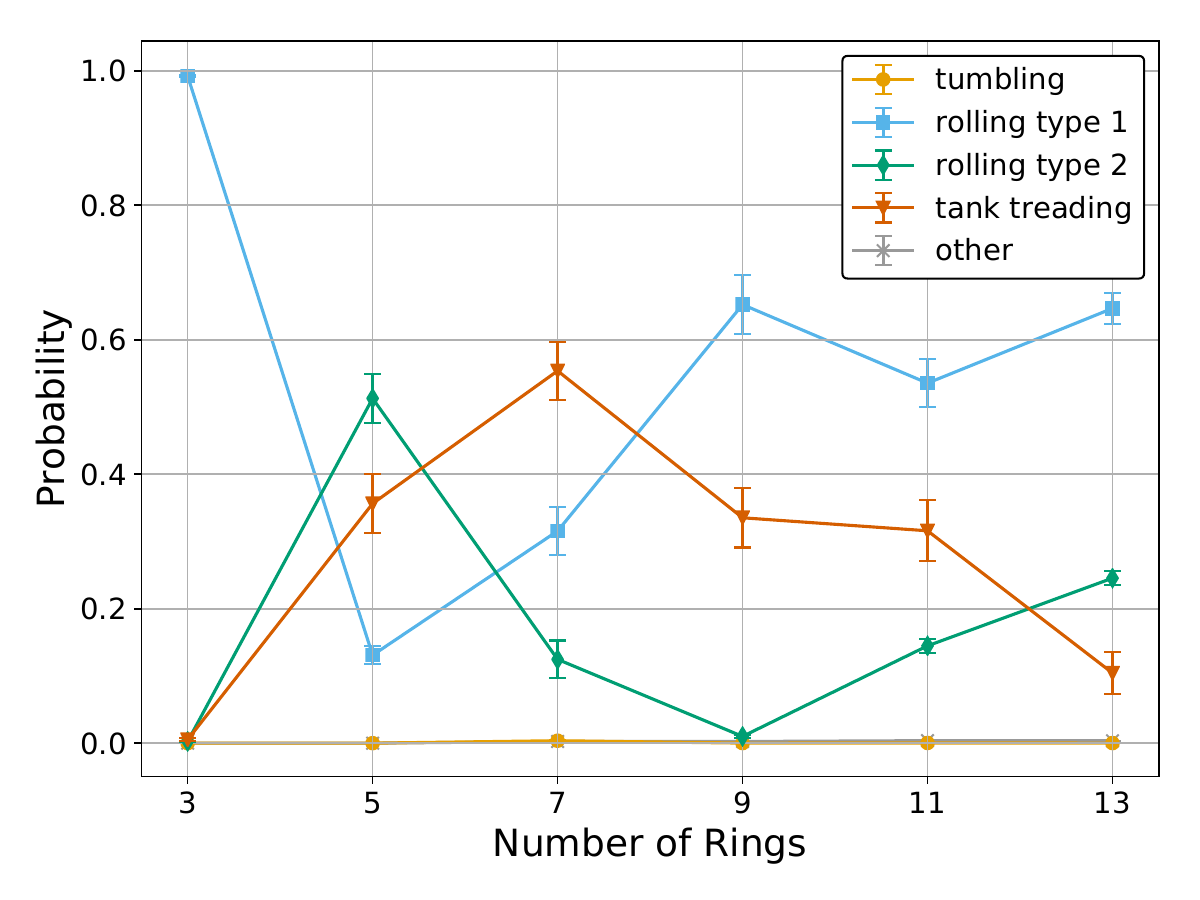}
    \includegraphics[width=0.49\textwidth]{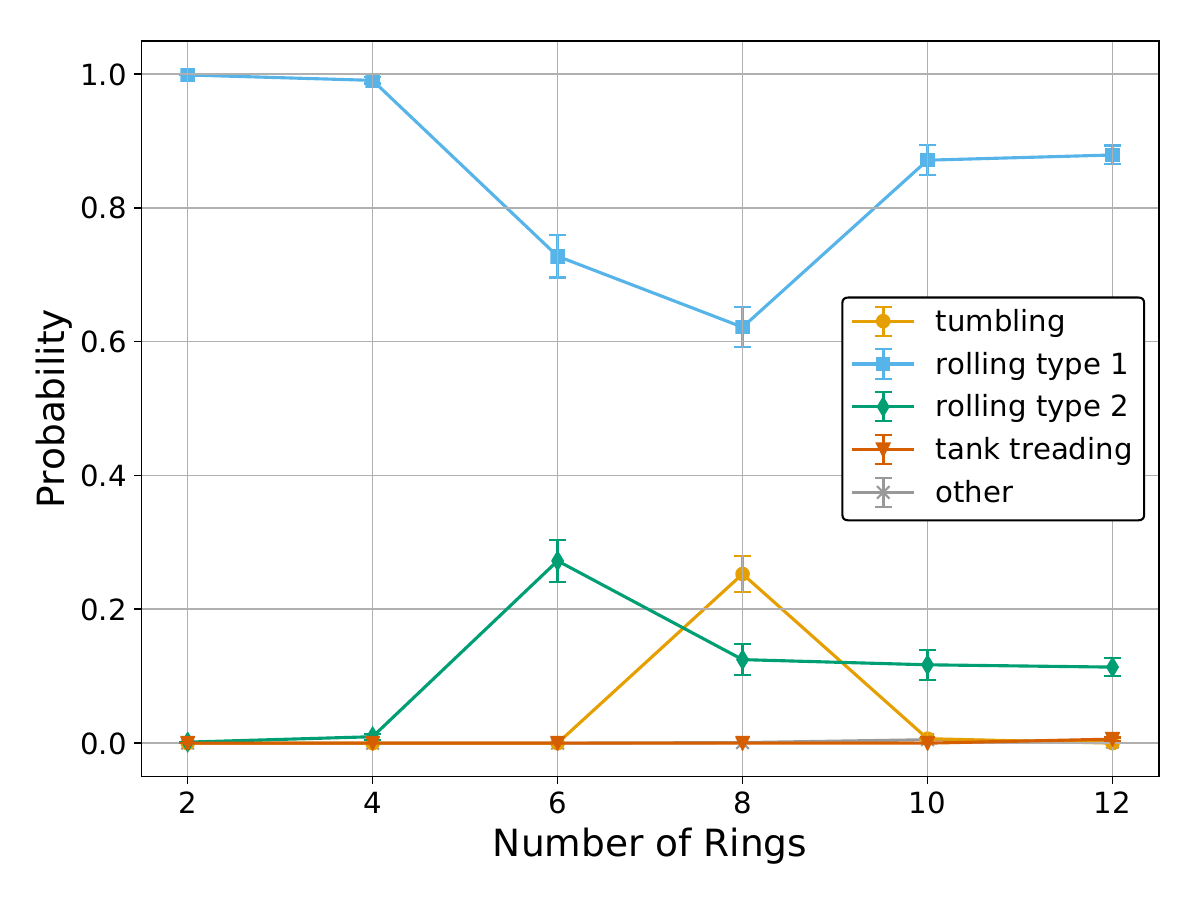}
  \caption{\justifying\small{ Probabilities of different dynamic behaviors in catenanes with an (a) odd and (b) even number of rings.
  For each [n]catenane, we ran 100 independent simulations. We calculated the probability of each behavior as the time spent in that mode relative to the total simulation time.
  }}
    \label{fig:modes}
\end{figure}

\begin{figure}
    \centering
    \includegraphics[width=0.49\textwidth]{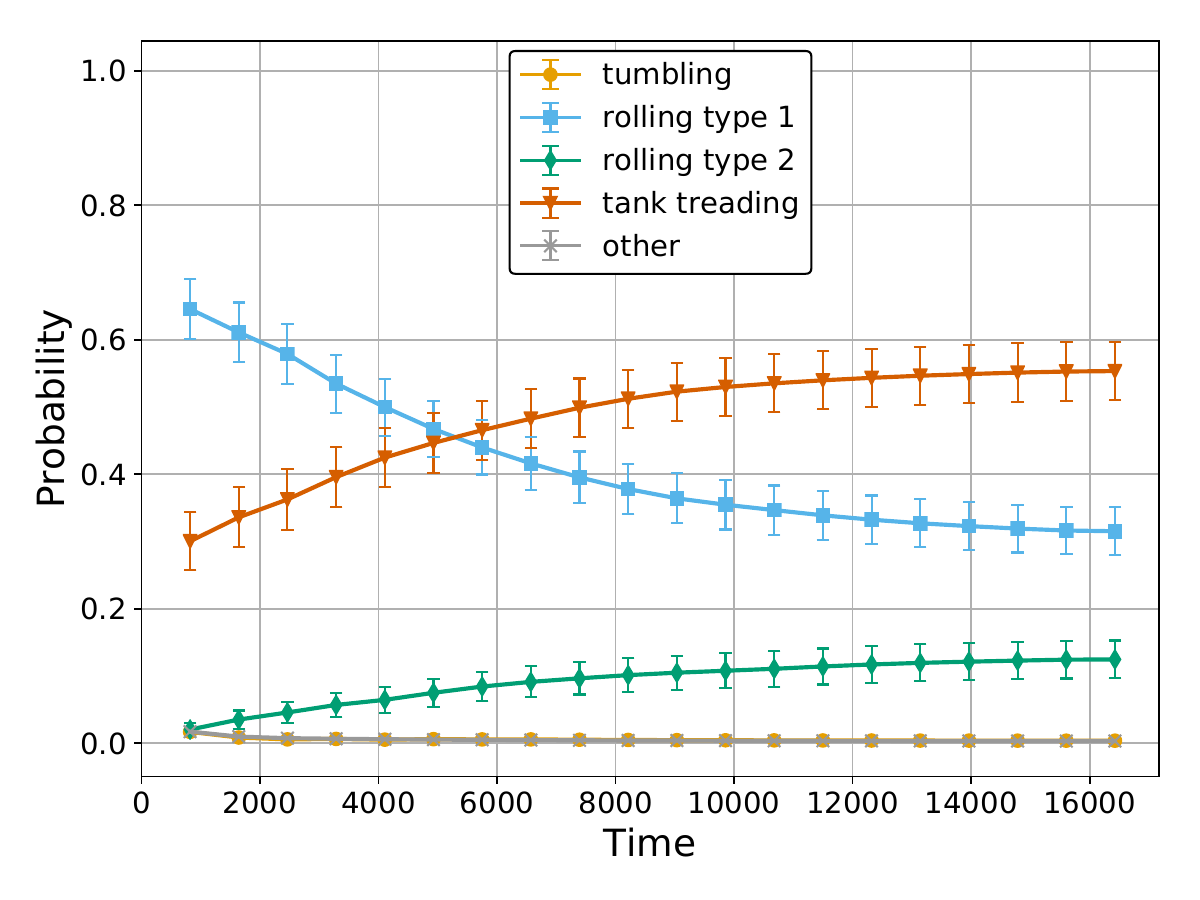}
  \caption{\justifying\small{ Time evolution of the cumulative probabilities of the different dynamic behaviors, calculated using data from the start of the simulations up to the indicated time. 
  Probabilities are averaged over 100 independent simulations, with error bars representing the associated uncertainty.}}
  \label{fig:modes_vs_time_7}
\end{figure}

At a high shear rate, polycatenanes exhibit several distinct dynamical behaviors, illustrated in Fig.~\ref{fig:intro}.
We classified these behaviors based on the components of the gyration tensor, which have unique average values in each mode as detailed in SM Table~\ref{tab:classification_thresholds}.
The four dominant behaviors are tumbling, two types of rolling, and tank-treading.
When tumbling, the polymer rotates and stretches end-over-end in the flow-gradient (\(xy\)) plane.
When rolling, the polymer is extended in the vorticity direction (\(z\)) and spins in the flow-gradient plane.
This behavior is consistent with the “mode I” dynamics reported by Kuei et al. for flexible bead–spring chains~\cite{kuei2015dynamics}.
We distinguished between rolling type 1, where the polymer is extended, and rolling type 2, where the polymer is compact.
The most striking behavior is tank-treading, where the polymer fully stretches in the flow direction and rings adopt a flattened profile.
%Cite Doyle and Huang when it comes out racetrack \Alex{CITE} configuration.
Rings alternate between being aligned in the flow-gradient plane, which rotate due to shear forces, and aligned in the flow-vorticity plane, which do not rotate at all.
This behavior is notable, because it results in an overall configuration that is extended and steady in a rotational flow field.
Individual components (every other ring) rotate, but the large-scale structure is irrotational.
See SM Video \texttt{7catenane\_TT.mp4} for a visualization of this tank-treading.
%Rings with a flow-gradient orientation rotate due to shear forces.
%Rings with a flow-vorticity orientation do not rotate at all.

We note that the rotation rate depends on the ring's location in the polymer.
The end rings rotate at $8.40 \pm 0.01$ revolutions per time unit, while the interior rings rotate slightly faster at $10.00 \pm 0.13$ revolutions per time unit.
We attribute this to the difference in ring extent in gradient direction. 
As evident from Fig.~\ref{fig:intro}b, the end rings are narrower in gradient (\(y\)) direction ($G_{yy}=1.94 \pm 0.02$) with a pinched configuration on one side when tank-treading.
The inner rings are symmetric and wider in gradient direction ($G_{yy}=2.50 \pm 0.01$), resulting in greater differences in hydrodynamic forces between the top and bottom of the ring and therefore faster rotations.

\begin{figure}
  \centering
  \begin{subfigure}[t]{0.48\textwidth}
    \centering
    \includegraphics[width=\textwidth]{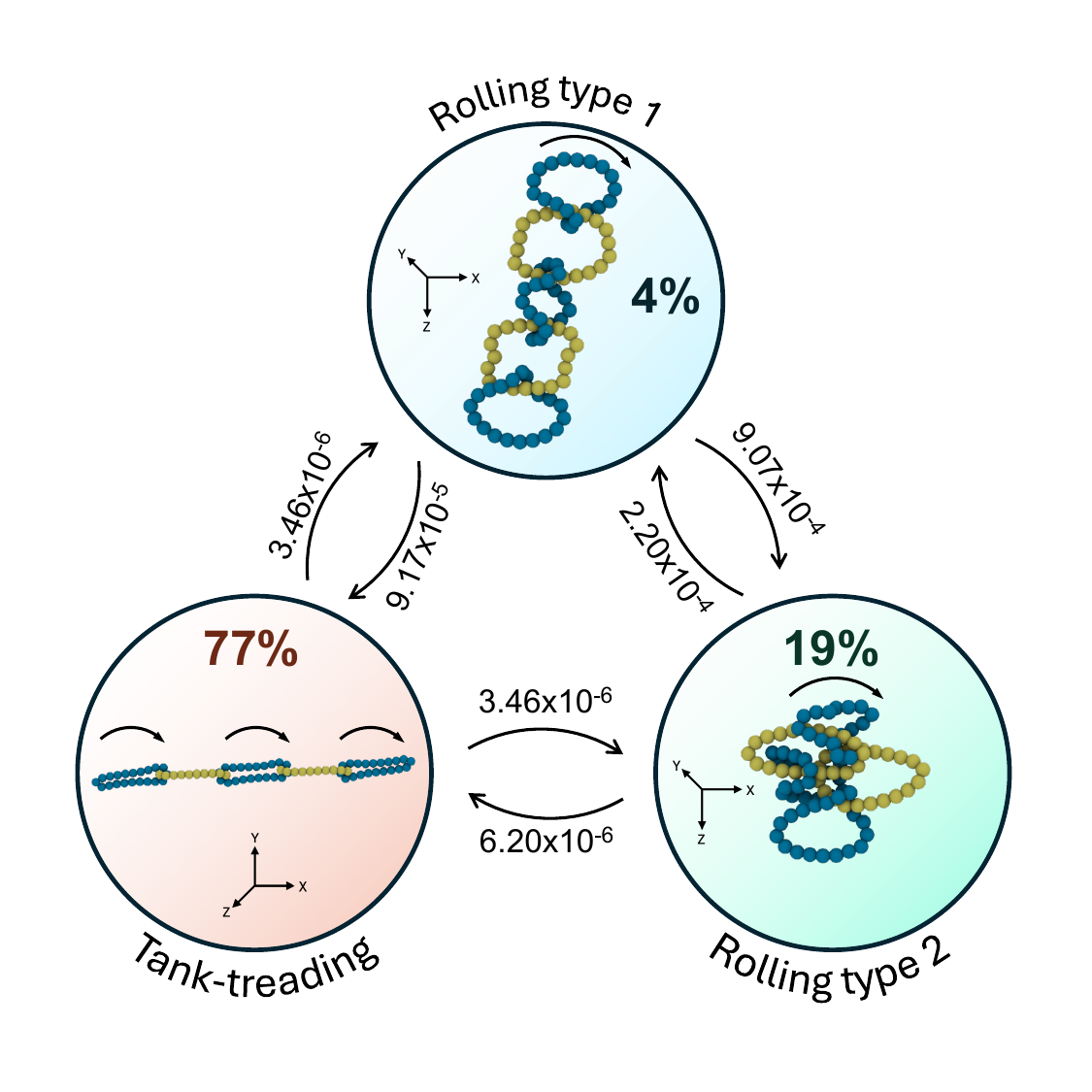}
  \end{subfigure}
   
  \caption{\justifying\small{ Illustration of the Markov state model (MSM) for [5]catenane.  Each dynamic behavior (rolling type 1, rolling type 2, tank-treading) is represented as a state, or node on a graph.
  Transition rates between these states are calculated from simulation and are represented as transitions between states, or directed edges on a graph.
  These rates are given as dimensionless frequencies.
  State probabilities are calculated from the MSM steady state and are given as a percentage.
  Note that [5]catenane does not exhibit tumbling, so it is excluded from this illustration.}}
  \label{fig:MSM}
\end{figure}

Remarkably, tank-treading is only observed in catenanes with an odd number of rings.
As demonstrated in Figs.~\ref{fig:modes}a and~b, all catenanes with 4 to 13 rings exhibit some degree of rolling type 1, rolling type 2, and tumbling, but only catenanes with odd numbers of rings demonstrate significant tank-treading.
For odd polycatenanes, we observed tank-treading most frequently with [7]catenane, where it occurs 55.4\% of the time.
We note that a [12]catenane tank-treads with a very low probability of 0.6\%, an interesting exception that we discuss later, but even polycatenanes mostly exhibit rolling behaviors in shear.
We reason that only odd polycatenanes tank-tread because both end rings need to be stably anchored.
A ring oriented in the flow-gradient plane can rotate internally in response to a shear force, but a ring oriented in the flow-vorticity plane needs to tumble or roll in response to a shear force.
In a stretched configuration, ring orientations must alternate orthogonally to minimize repulsive clashes at their interlocked interfaces.
This alternating configuration of rings is observed when polycatenanes are stretched in extensional flow~\cite{seyedi2026mechanically}.
Because of this, odd polycatenanes can have end rings with the same orientation, but even polycatenanes cannot.
In odd polycatenanes, with end rings oriented in the same flow-gradient plane and rotating internally, these end rings can serve as stable anchors for a completely elongated chain.
In even polycatenanes, with one end ring necessarily oriented in the flow-vorticity plane, it cannot rotate internally and must roll or tumble in response to a shear force, which disturbs the stretched configuration and makes elongated tank-treading unstable.

To ensure that tank-treading is a natural behavior and not an artifact of the initial conditions, we started all simulations from configurations sampled from the equilibrium (no shear) distribution.
These equilibrium configurations are not extended or oriented to promote tank-treading.
To further demonstrate the stability of tank-treading, in Fig.~\ref{fig:modes_vs_time_7} we show that its probability actually increases as a simulation progresses.
In this example, for a [7]catenane we see that the initial probability of observing tank-treading is 30\%, but this increases to 55.4\% on our simulation time scale.  
Data for other odd polycatenanes is present in SM Fig.~\ref{fig:modes_vs_time_odd} and shows similar trends, with the probability of tank-treading either increasing or stable on simulation time scales.

\onecolumngrid
\begin{center}
\begin{table}[h]
\centering
\caption{Probabilities of dynamic behaviors of odd polycatenanes observed directly from our simulations (left) and calculated from the steady state of a Markov model (right).}
\label{tab:probs}
\renewcommand{\arraystretch}{1.1}
\setlength{\tabcolsep}{4pt}
\begin{tabular}{ccccccccccc}
& \multicolumn{2}{c}{Tumbling} & \multicolumn{2}{c}{Rolling Type 1} & \multicolumn{2}{c}{Rolling Type 2} & \multicolumn{2}{c}{tank-treading} & \multicolumn{2}{c}{Other} \\
Polymer & Simulation & MSM & Simulation & MSM & Simulation & MSM & Simulation & MSM & Simulation & MSM \\
\toprule
\textup{[3]}catenane & 0.0000 & 0.0000 & 0.9922 & 1.0000 & 0.0026 & 0.0000 & 0.0052 & 0.0000 & 0.0000 & 0.0000 \\
\midrule
\textup{[5]}catenane & 0.0000 & 0.0000 & 0.1308 & 0.0437 & 0.5128 & 0.1892 & 0.3564 & 0.7671 & 0.0000 & 0.0000 \\
\midrule
\textup{[7]}catenane & 0.0036 & 0.0000 & 0.3155 & 0.0000 & 0.1246 & 0.0000 & 0.5537 & 1.0000 & 0.0025 & 0.0000 \\
\midrule
\textup{[9]}catenane & 0.0001 & 0.0000 & 0.6522 & 0.0000 & 0.0100 & 0.0000 & 0.3352 & 1.0000 & 0.0025 & 0.0000 \\
\midrule
\textup{[11]}catenane & 0.0003 & 0.0000 & 0.5354 & 0.0000 & 0.1447 & 0.0000 & 0.3158 & 1.0000 & 0.0037 & 0.0000 \\
\midrule
\textup{[13]}catenane & 0.0001 & 0.0000 & 0.6466 & 0.0000 & 0.2454 & 0.0000 & 0.1043 & 1.0000 & 0.0036 & 0.0000 \\
\bottomrule

\end{tabular}
\end{table}
\end{center}
\twocolumngrid

To estimate the stability of tank-treading on longer time scales, we employed a Markov model.
We represented each dynamical behavior as a state in a continuous-time Markov-state model (MSM) \cite{albaugh2022simulating} and estimated the transition rates between states from the simulation trajectories. 
The transition rate from one state \(i\) to another \(j\) was calculated as

\begin{equation}
k_{ij} = \frac{N_{ij}}{T_i},
\end{equation}

where \(N_{ij}\) is the total number of observed transitions from \(i\) to \(j\), and \(T_i\) is the total time spent in state \(i\). 
These rates were assembled into a generator matrix \(\mathbf{Q}\), whose elements are defined as

\begin{equation}
Q_{ij} =
\begin{cases}
k_{ij}, & i \neq j, \\
-\sum\limits_{j \neq i} k_{ij}, & i = j.
\end{cases}
\end{equation}
The steady-state population vector, \(\boldsymbol{\pi}\), was obtained by solving \(\mathbf{Q}^{T}\boldsymbol{\pi}=0\) subject to the normalization condition \(\sum_i \pi_i = 1\).
Thus, the steady-state populations correspond to the normalized eigenvector of \(\mathbf{Q}^{T}\) associated with the zero (largest) eigenvalue. 
The MSM for a [5]catenane is illustrated in Fig.~\ref{fig:MSM}, where each behavior is a state (node) and transition rates estimated from our simulations connect these states (directed edges).  
The eigenvector analysis of the generator matrix implies that the steady state populations for this [5]catenane are 4\% rolling type 1, 19\% rolling type 2, and 77\% tank-treading.
The 77\% tank-treading predicted by the Markov model is greater than the 36\% tank-treading that we observe from simulations, implying that tank-treading is even more stable on long time scales than our simulations suggest.
This trend is observed across all odd polycatenanes with significant tank-treading ([5] through [13]), as shown in Table~\ref{tab:probs}.
Data for even polycatenanes is given in SM Table~\ref{tab:even_catenantes}.
In fact for [7]catenane to [13]catenane the MSM predicts that the long-time probability of tank-treading is 100\%.
This occurs because in those systems we never observe a transition to another behavior after the system starts tank-treading.
As a result, the rate of leaving a tank-treading state is 0 and these states are complete sinks in the Markov model.
This implies that if we were able to simulate long enough, every polymer would eventually transition into a tank-treading behavior and never stop.
The results are further evidence that tank-treading is a highly stable behavior in odd polycatenanes.

\begin{figure}
  \centering
  \begin{subfigure}[t]{0.49\textwidth}
    \centering
    \includegraphics[width=\textwidth]{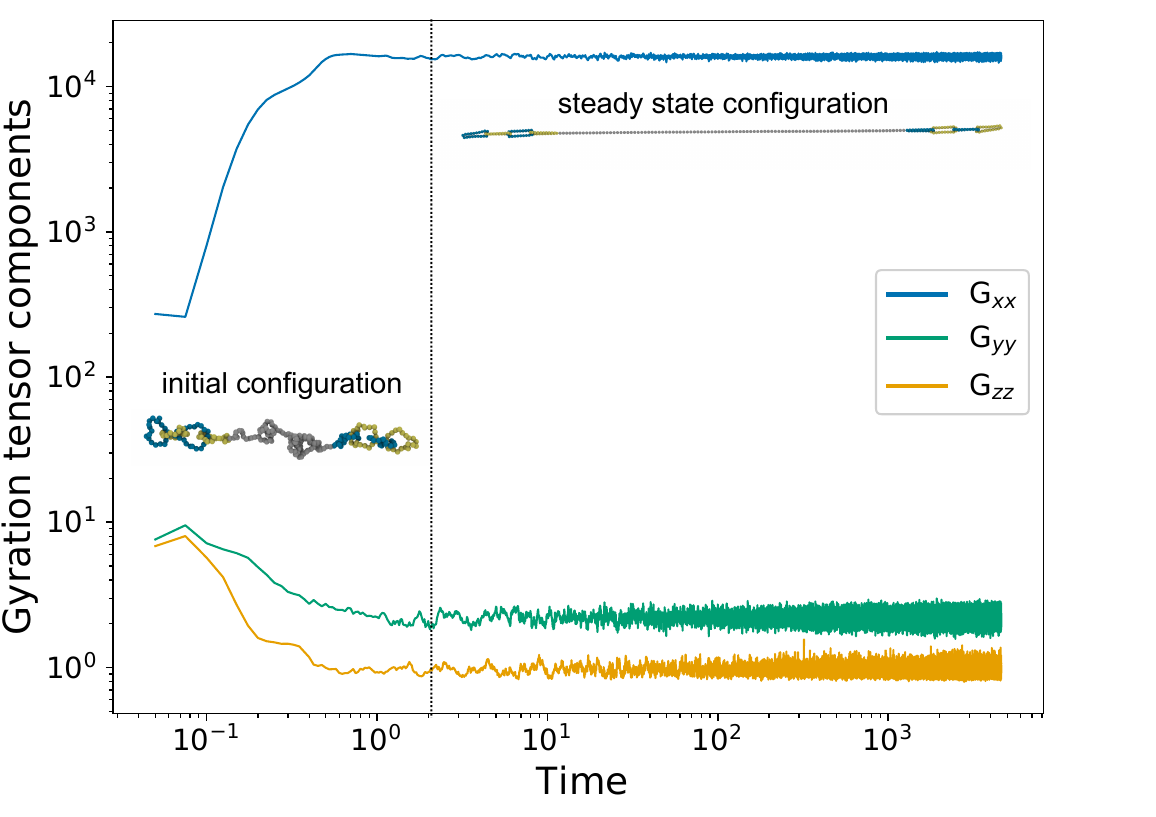}
    \label{fig:gyration_tensor_linear}
  \end{subfigure}
  \caption{\justifying\small{ Gyration tensor components as a function of time for a linear polymer (\(N=100\)) attached to a [4]catenane on each side, with initial and steady state trajectory snapshots.
  In the initial configuration, the linear polymer is in a compact, random configuration.
  At steady state, the linear polymer is completely elongated and catenane rings in the flow-gradient plane rotate in a tank-treading behavior.
  The logarithmic time axis highlights that the random initial configuration quickly stabilizes to the stretched state.}}
  \label{fig:linear_stretch}
\end{figure}

Our simulations further suggest that we can use odd polycatenane tank-treading to stretch other molecules in strong shear.
As demonstrated in SM Fig.~\ref{SI/gyration_linear}, a linear polymer undergoes continuous stretch-tumble cycles in shear flow and does not adopt a stable conformation.
If we attach a [4]catenane to each end of this linear polymer, the combined polymer will now completely and stably stretch in that same shear flow, shown in Fig.~\ref{fig:linear_stretch}, resulting in a persistent elongated configuration.
We used an even number of rings in each catenane section so that its two ends take opposite orientations. 
The outer ring sits in the flow-gradient plane and rotates, acting as the hydrodynamic anchor that holds the chain extended, while the inner ring sits in the flow-vorticity plane and does not rotate. 
Bonding the linear polymer to this fixed inner ring gives a stable attachment point that does not rotate; an odd section would instead place a rotating ring at the connection point.

We show that even from a starting configuration where the linear polymer is in random, compact state, the catenane sections can quickly stretch the entire combined polymer into a stable elongated state.
This proof-of-concept demonstrates the possibility of using catenanes to stretch a variety of molecules in rotational flows.

We attribute the rare tank-treading behavior of [12]catenane from Fig.~\ref{fig:modes}b to this ability of catenanes to stretch a molecule between them.
There is probability of 0.6\% of observing tank-treading in [12]catenane, which we show in SM Video \texttt{12catenane\_TT.mp4}.
As shown in SM Fig.~\ref{SI/12_rings}, when tank-treading, the outermost rings of [12]catenane can both orient in the flow-gradient plane, like odd-polycatenanes when tank-treading.
The outer 3 rings on either side of [12]catenane adopt alternating flow-gradient and flow-vorticity orientations, with the flow-gradient rings rotating.
The inner 6 rings, however, twist out of the flow-gradient and flow-vorticity planes.
With the [12]catenane, then, there are sufficient rings to twist enough in the middle to support tank-treading orientations on either end.
In this way, the rare tank-treading observed in this even catenane is a case of two tank-treading [3]catenane ends stretching a [6]catenane motif in the middle.

\section{Discussion}
Our results hint at several possible applications for this tank-treading behavior.
High molecular weight polymers are known to reduce drag in pipe flow~\cite{larson2003analysis,perkins1997single,dimitropoulos1998direct,paterson1970turbulent}, at least partially due to the suppression of vortex formation caused by extended polymer conformations.
Tank-treading results in persistent extended polymer configurations in a rotational flow, which may result in better drag reducing behavior than typical linear polymers, which stretch and tumble through un-extended configurations.
Additionally, elongated polymers added to liquid fuels will suppress the formation of small droplets.
By forming larger droplets, fuel is less likely to ignite during an accidental release, increasing safety in automobiles and aircraft~\cite{chao1984antimisting,wei2015megasupramolecules,lhota2024mist}.
The extended tank-treading configurations of polycatenanes may prove beneficial in this application, as well.
Our results in Fig.~\ref{fig:linear_stretch} directly show that attaching catenanes to either end of a molecule can stretch that molecule in shear flow.
This may have potential biological or microfluidic applications.
For instance, one may be able to unfold a protein or unwind DNA in a shear or other rotational flow, or stretch molecules for imaging and measurement in microfluidic channels.

While interesting, these potential applications and our results will need to be evaluated in increasingly realistic systems and models.
The Weissenberg number (\(\mathrm{Wi=\tau_{R}\dot{\gamma}}\), where \(\tau_{\mathrm{R}}\) is the longest relaxation time in the polymer) gives the ratio of deformation rate to the intrinsic time scale of relaxation for a polymer.
Our simulations of [5]-, [7]-, [9]-, [11]-, and [13]catenane imply Wi of \(2.7\times10^4\), \(4.8\times10^4\), \(7.1\times10^4\), \(1.1\times10^5\), and \(1.4\times10^5\), respectively.
While mechanical interlocks are estimated to be at least as strong as chemical bonds~\cite{lee2016mechanical}, it is unclear if a catenane could realistically maintain integrity in such strong flows.
Future work will therefore focus on developing models to investigate degradation at high shear rates, using more realistic atomistic models to confirm the presence of tank-treading, and exploring a wide range of conditions and polymer configurations to discover tank-treading at milder shear rates.

\section{Conclusions}
In this work, we revealed a unique flow-induced dynamic behavior in linear polycatenanes subject to strong shear flow using coarse-grained Brownian dynamics simulations with excluded volume and hydrodynamic interactions. 
Unlike conventional polymers that tumble under shear, polycatenanes show stable tank-treading dynamics characterized by rotation of individual rings while the polymer maintains an extended overall configuration aligned with the flow. 
This extended state, which resembles conformations typically observed under extensional flow, emerges despite the rotational component in shear flow.

We demonstrated that tank-treading, with rare exception, is exclusive to odd polycatenanes while rolling is the prominent dynamics in even polycatenanes.
Furthermore, our Markov state model confirms that the tank-treading state represents a stable dynamic behavior rather than a transient state. 
Indeed, the MSM showed that tank-treading is the only steady state behavior of odd polycatenanes with 7 or more rings.
Finally, we demonstrated the possibility of using tank-treading polycatenanes as molecular actuators, where they enable the extension of other molecules in strong shear flow.

\section*{Acknowledgments}
The authors gratefully acknowledge the financial support of Wayne State University and the computational resources of Wayne State University's High Performance Computing.

\section*{Supplemental Materials}
The Supplemental Materials contains details of the computational methodology, plots of the gyration tensor components and their eigenvalues, a trajectory snapshot of the \textup{[12]}catenane, the time evolution of cumulative probabilities for different dynamic behaviors of odd polycatenanes, gyration tensor components of a linear polymer under strong shear flow, tables summarizing the probabilities of dynamic behaviors for even polycatenanes, classification thresholds used to identify dynamic modes, and videos of the observed dynamic behaviors.

\section*{Data Availability}
Data will be available in a persistent Zenodo repository. 
Simulation code is available at \href{https://github.com/albaugh/polymer}{https://github.com/albaugh/polymer}.

%\clearpage
\bibliography{biblio.bib}

\clearpage

\onecolumngrid
\appendix
\setcounter{figure}{0}
\setcounter{table}{0}  
\renewcommand{\thefigure}{\Alph{section}.\arabic{figure}}  
\renewcommand{\thetable}{\Alph{section}.\arabic{table}}   

\section{}
%\section{Simulation Details}

Polymers consist of \(N\) beads interacting through bonded and non-bonded potentials. 
The pairwise non-bonded interaction is the Weeks--Chandler--Andersen potential~\cite{weeks1971role}
\begin{equation}
U_{\mathrm{WCA}} = \sum_{\substack{i,j \\ j>i}} \left\{
\begin{array}{ll}
4\epsilon \left[ \left(\frac{\sigma}{r_{ij}} \right)^{12} - \left(\frac{\sigma}{r_{ij}} \right)^{6} \right] - U_c, & r_{ij} < r_c \\[10pt]
0, & r_{ij} \geq r_c
\end{array}
\right\},
\end{equation}
with \(r_c=2.2449241\sigma\), \(U_c = 4 \epsilon \left[ \left(\frac{\sigma}{r_c} \right)^{12} - \left(\frac{\sigma}{r_c} \right)^{6} \right]\), and \(\epsilon=100 k_{B}T\).

Bonds are modeled with FENE springs~\cite{kremer1990dynamics}:
\begin{equation}
U_{\mathrm{FENE}} = -\sum_{i,j} \frac{1}{2} k_s r_{\text{max}}^2 \log \left[1 - \left(\frac{r_{ij}}{r_{\text{max}}} \right)^2 \right], \quad r_{ij} < r_{\text{max}},
\end{equation}
where \(k_{s}=30\epsilon/\sigma^2\) and \(r_{\mathrm{max}}=1.2\sigma\)
%\Ali{SEE COMMENT LINE IN LATEX CODE. 
%from polymer code:"ff->k["BEAD-BEAD"] = eps_over_kBT * k_sig2_over_eps;" 
%we use eps\_over\_kbt=100. so the code also scales up the spring stiffness so the actual $k_s$ will be \(k_{s}=3000\varepsilon/\sigma^2\) not \(k_{s}=30\varepsilon/\sigma^2\)}
These parameters prevent bond crossing, improve numerical stability at high shear rates, and result in a stiffer polymer than the standard Kremer-Grest paramaterization (\(\epsilon =k_{B}T\), \(r_{\mathrm{max}}=1.5\sigma\)).
For linear segments, \(r_{\mathrm{max}}=1.5\sigma\). 
The total potential is \(U = U_{\text{WCA}} + U_{\text{FENE}}\) and \(\mathbf{f} = -\nabla U\).

The diffusivity matrix incorporates hydrodynamic interactions:
\begin{equation}
\mathbf{D}_{ij} = \frac{k_B T}{\zeta} \left[ (1 - \delta_{ij}) \bm{\Omega}_{ij} + \delta_{ij} \mathbf{I} \right],
\end{equation}
with the Rotne--Prager--Yamakawa tensor \(\bm{\Omega}_{ij}\)~\cite{rotne1969variational,yamakawa1970transport}:
\begin{equation}
\bm{\Omega}_{ij} = 
\begin{cases} 
\frac{3a}{4r_{ij}} 
\left[ 
\left( 1 + \frac{2a^2}{3r_{ij}^2} \right) \mathbf{I} + 
\left( 1 - \frac{2a^2}{r_{ij}^2} \right) \frac{\mathbf{r}_{ij} \mathbf{r}_{ij}}{r_{ij}^2}
\right], & r_{ij} \geq 2a, \\[15pt]
\left( 1 - \frac{9r_{ij}}{32a} \right) \mathbf{I} + 
\left( \frac{3r_{ij}}{32a} \right)\frac{\mathbf{r}_{ij} \mathbf{r}_{ij}}{r_{ij}^2} , & r_{ij} < 2a,
\end{cases}
\label{eq:RPY tensor}
\end{equation}

where $\zeta=6\pi\eta_sa$ is the drag coefficient, $\eta_s$ is the solvent viscosity, and 
\(a=\sigma/2\) is the hydrodynamic radius of the beads.
The solvent velocity \(\mathbf{v}_{i}\) is zero at equilibrium and \([\dot{\gamma} (y_i-y_{\mathrm{COM}}), 0, 0]^\top\) in shear flow. 
Relaxation times are \(\tau_R = R_g^2 / 6 D_{\mathrm{COM}}\) calculated using the Kirkwood formalism~\cite{dunweg2002corrections}.

Variables are nondimensionalized with length unit \(a\), energy unit \(k_B T\), and time unit \(\tau = \zeta a^2 / k_B T\). The Semi-implicit scheme yields
\begin{align}
\mathbf{r}^*(t^*{+}\Delta t^*) &= \mathbf{r}^*(t^*) + \sqrt{2\Delta t^*}\, \mathbf{B}^*(t^*) \cdot \Delta \mathbf{w}^* \notag \\
&\quad + \Big[\mathbf{v}^*(t^*{+}\Delta t^*) + \mathbf{D}^*(t^*) \cdot \mathbf{f}^*(t^*{+}\Delta t^*)\Big] \Delta t^*,
\end{align}
solved iteratively via Jacobian-free Newton-Krylov through the NITSOL software ~\cite{pernice1998nitsol}. Initial configurations use equilibrium bond lengths followed by energy minimization. Simulations run for (\(4.9 \times 10^7\) -- \(3.5 \times 10^8\) steps, \(\Delta t^* = 10^{-4}\)).

\iffalse
The polymer contribution to the stress tensor is

\begin{equation}
\frac{\boldsymbol{\tau}^p}{n_p k_B T}
=
-(N - 1)\mathbf{I}
-
\sum_{i=1}^{N-1}
\sum_{j=i+1}^{N}
\mathbf{f}^*_{ij}\mathbf{r}^*_{ij},
\label{eq:stress_tensor}
\end{equation}

from which \(\tau_{xy}\) and \(N_1 = \tau^p_{xx} - \tau^p_{yy}\) are obtained~\cite{bird1977dynamics,Graham_2018,bosko2008effect}.
\fi

\begin{figure}[p]
    \centering    \includegraphics[width=\textwidth,keepaspectratio]{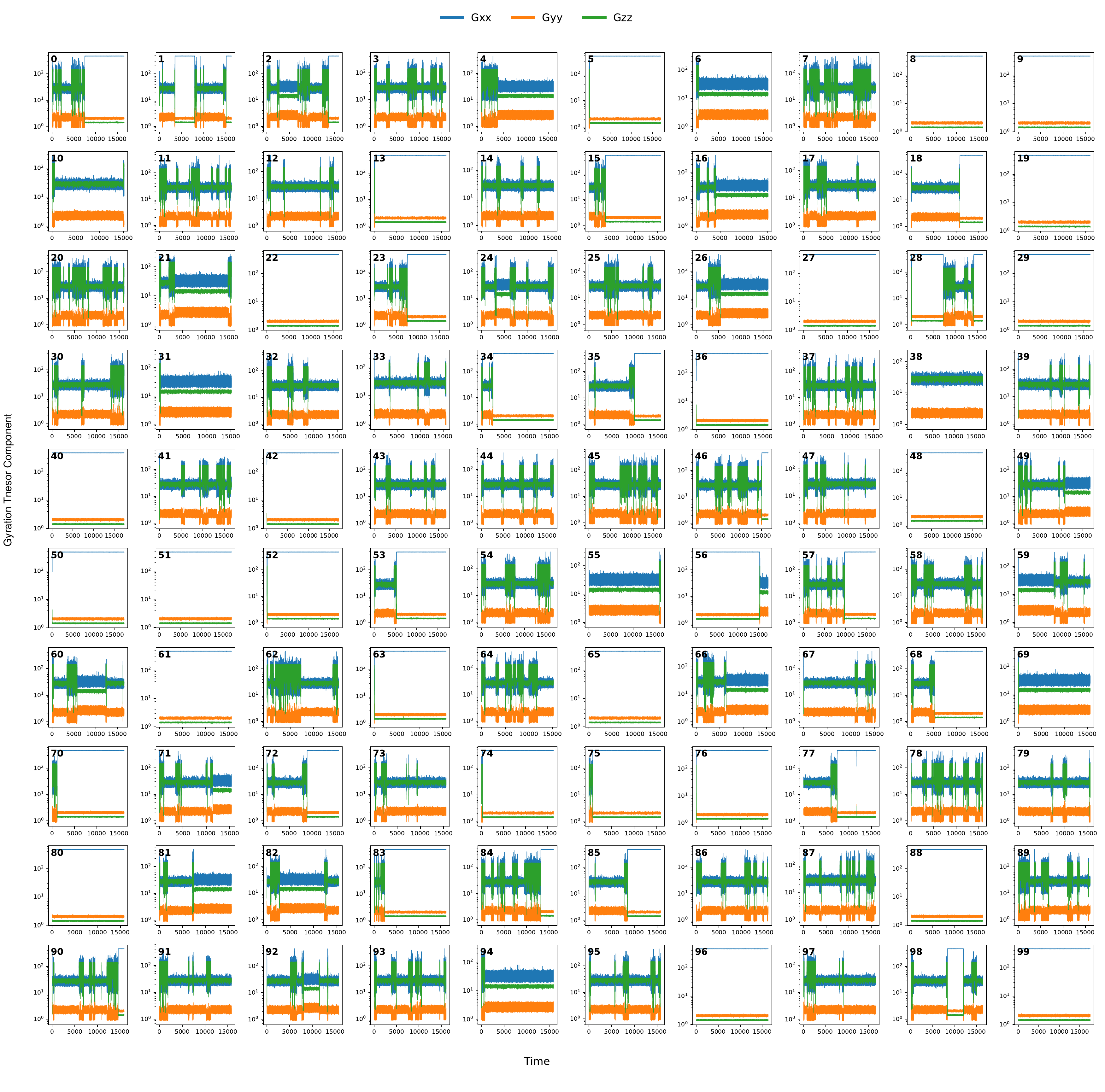}
    \caption{Gyration tensor components for [5]catenane.  Each graph is time series data for the individual components (\(G_{xx}\), \(G_{yy}\), \(G_{zz}\)) for each of 100 independent simulations.}
    \label{SI/rg}
\end{figure}

\begin{figure}[p]
    \centering    \includegraphics[width=\textwidth,keepaspectratio]{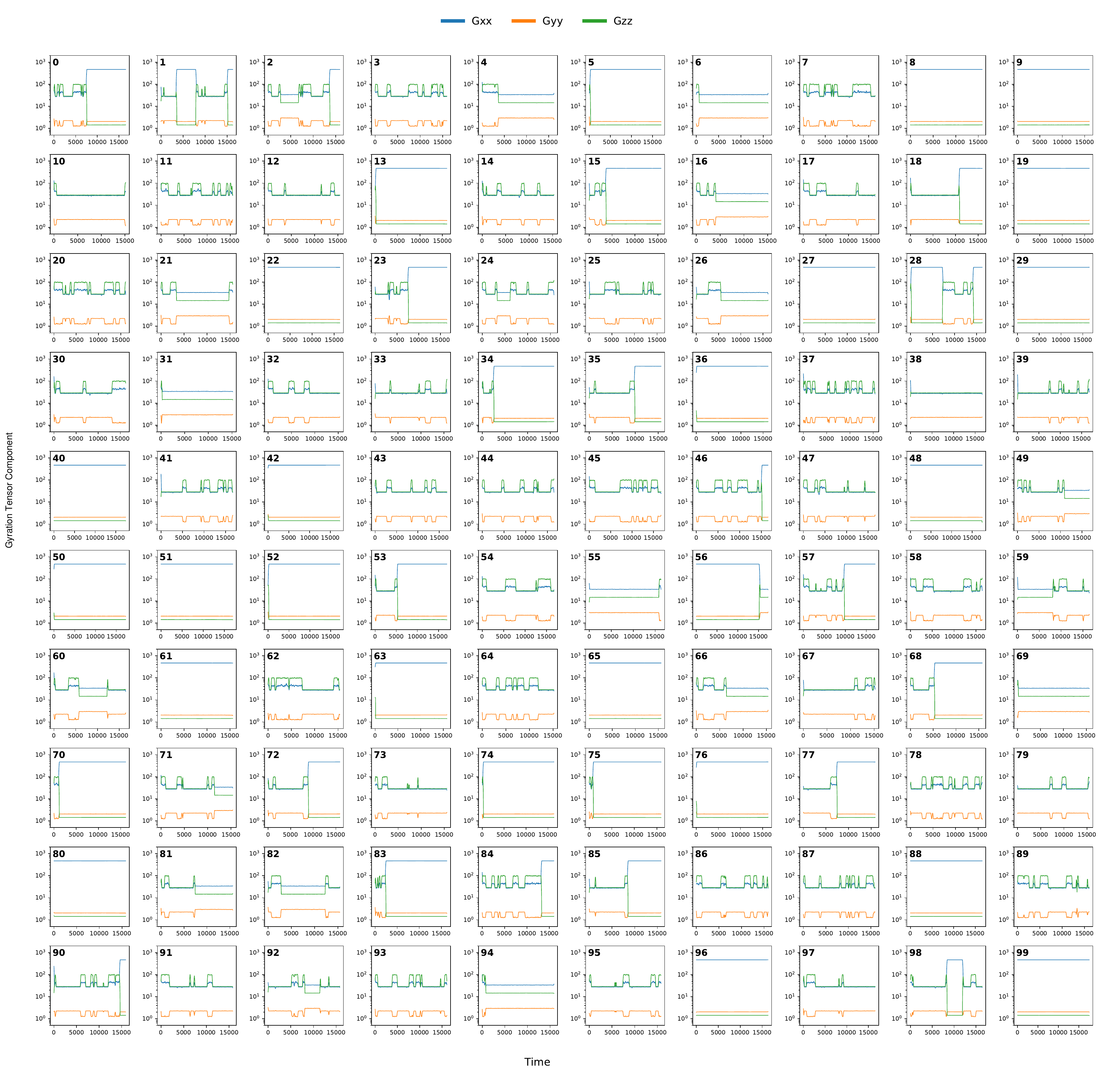}
    \caption{Gyration tensor components for [5]catenane after applying a moving time average filter.  Each graph is time series data for the individual components (\(G_{xx}\), \(G_{yy}\), \(G_{zz}\)) for each of 100 independent simulations.}
    \label{SI/rg_smooth}
\end{figure}

\begin{figure}[p]
    \centering    \includegraphics[width=\textwidth,keepaspectratio]{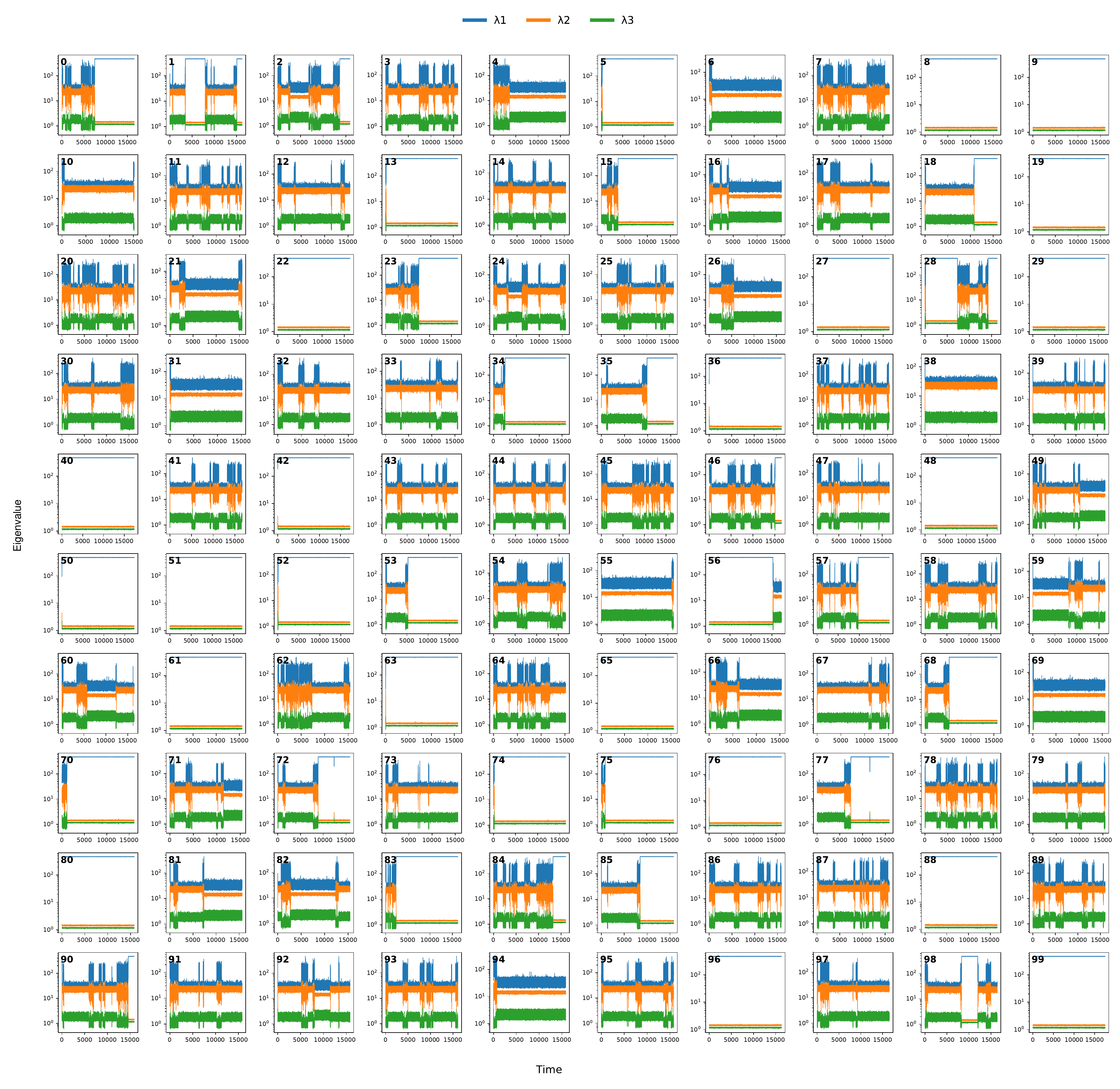}
    \caption{Eigenvalues of the gyration tensor for [5]catenane.  Each graph is time series data for the eigenvalues \( \lambda_{1 }\), \( \lambda_{2} \), \( \lambda_{3} \)) for each of 100 independent simulations.}
    \label{SI/eig}
\end{figure}

The dynamic behaviors are characterized using the gyration tensor components
\begin{equation}
G_{\alpha\beta}
=
\frac{1}{N}
\sum_{i=1}^{N}
(r_{i,\alpha}-r_{\mathrm{COM},\alpha})
(r_{i,\beta}-r_{\mathrm{COM},\beta}),
\qquad
\alpha,\beta \in \{x,y,z\},
\end{equation}
where \(r_{i,\alpha}\) is the \(\alpha\)-component of the position vector of bead \(i\), and \(r_{\mathrm{COM},\alpha}\) is the corresponding component of the center-of-mass position,
\begin{equation}
r_{\mathrm{COM},\alpha}
=
\frac{1}{N}
\sum_{i=1}^{N} r_{i,\alpha}.
\end{equation}
To reduce excessive noise in gyration tensor data arising from the thermal fluctuations, the components (\(G_{xx}\), \(G_{yy}\), \(G_{zz}\)) were smoothed prior to visualization and analysis using a centered moving-average filter. 
For each trajectory, an averaging window corresponding to approximately 1\% of the total number of data points was applied to each gyration tensor component, while the time coordinate was left unchanged. 
Figs.~\ref{SI/rg} and~\ref{SI/rg_smooth} show the original components before and after applying the filter, respectively, for an example [5]catenane.
For comparison, we also present the eigenvalues of the gyration tensor for the same system in Fig.~\ref{SI/eig}, which convey approximately equivalent information.
From Fig.~\ref{SI/rg_smooth} it is clear that the polycatenane has distinct regimes of behavior.
We classify these regimes based on the gyration tensor components according to Table~\ref{tab:classification_thresholds}.

\begin{table*}[h]
\centering
\caption{\justifying Classification thresholds used to identify dynamic modes in [n]catenanes. In all cases, tumbling and tank-treading require $G_{xx}>G_{yy}>G_{zz}$ and rolling requires $G_{zz}>G_{xx}>G_{yy}$. Rolling type~2 is assigned when the rolling ordering is satisfied but the rolling type~1 threshold on $G_{zz}$ is not.}
\label{tab:classification_thresholds}
\renewcommand{\arraystretch}{1.2}
\setlength{\tabcolsep}{15pt}
\begin{tabular}{cccccc}
Polymer & Tumbling & Rolling type 1 & Tank treading \\
\toprule
$[2]$catenane  & $G_{xx}<140,\ G_{yy} < 6 $  & $11 < G_{zz} < 31$         & $G_{zz}<2$ \\
$[3]$catenane  & $G_{xx}<140,\ G_{yy} < 6$   & $31 < G_{zz} < 61$         & $G_{zz}<2$ \\
$[4]$catenane  & $G_{xx}<400,\ G_{yy} < 6$   & $60 < G_{zz} < 90$         & $G_{zz}<2$ \\
$[5]$catenane  & $G_{xx}<400,\ G_{yy} < 6$   & $85 < G_{zz} < 115$        & $G_{zz}<2$ \\
$[6]$catenane  & $G_{xx}<400,\ G_{yy} < 6$   & $115 < G_{zz} < 145$       & $G_{zz}<2$ \\
$[7]$catenane  & $G_{xx}<800,\ G_{yy} < 8$   & $115 < G_{zz} < 185$       & $G_{zz}<2$ \\
$[8]$catenane  & $G_{xx}<600,\ 3 < G_{yy} < 11$   & $175 < G_{zz} < 235$    & $G_{zz}<2$ \\
$[9]$catenane  & $G_{xx}<800,\ G_{yy} < 8$   & $180 < G_{zz} < 300$         & $G_{zz}<2$ \\
$[10]$catenane & $G_{xx}<1200,\ 3 < G_{yy} < 11$  & $240 < G_{zz} < 360$    & $G_{yy}<2,\ G_{zz}<2$ \\
$[11]$catenane & $G_{xx}<1400,\ G_{yy} < 8$  & $240 < G_{zz} < 360$         & $G_{yy}<3,\ G_{zz}<3$ \\
$[12]$catenane & $G_{xx}<1600,\ 3 < G_{yy} < 11$  & $300 < G_{zz} < 460$    & $G_{yy}<3,\ G_{zz}<3$ \\
$[13]$catenane & $G_{xx}<2900,\ G_{yy} < 8$  & $350 < G_{zz} < 470$         & $G_{yy}<3,\ G_{zz}<3$ \\
\bottomrule
\end{tabular}
\end{table*}

\begin{figure*}[t]
\centering

\begin{subfigure}[t]{0.48\textwidth}
    \centering
    \begin{overpic}[width=\textwidth]{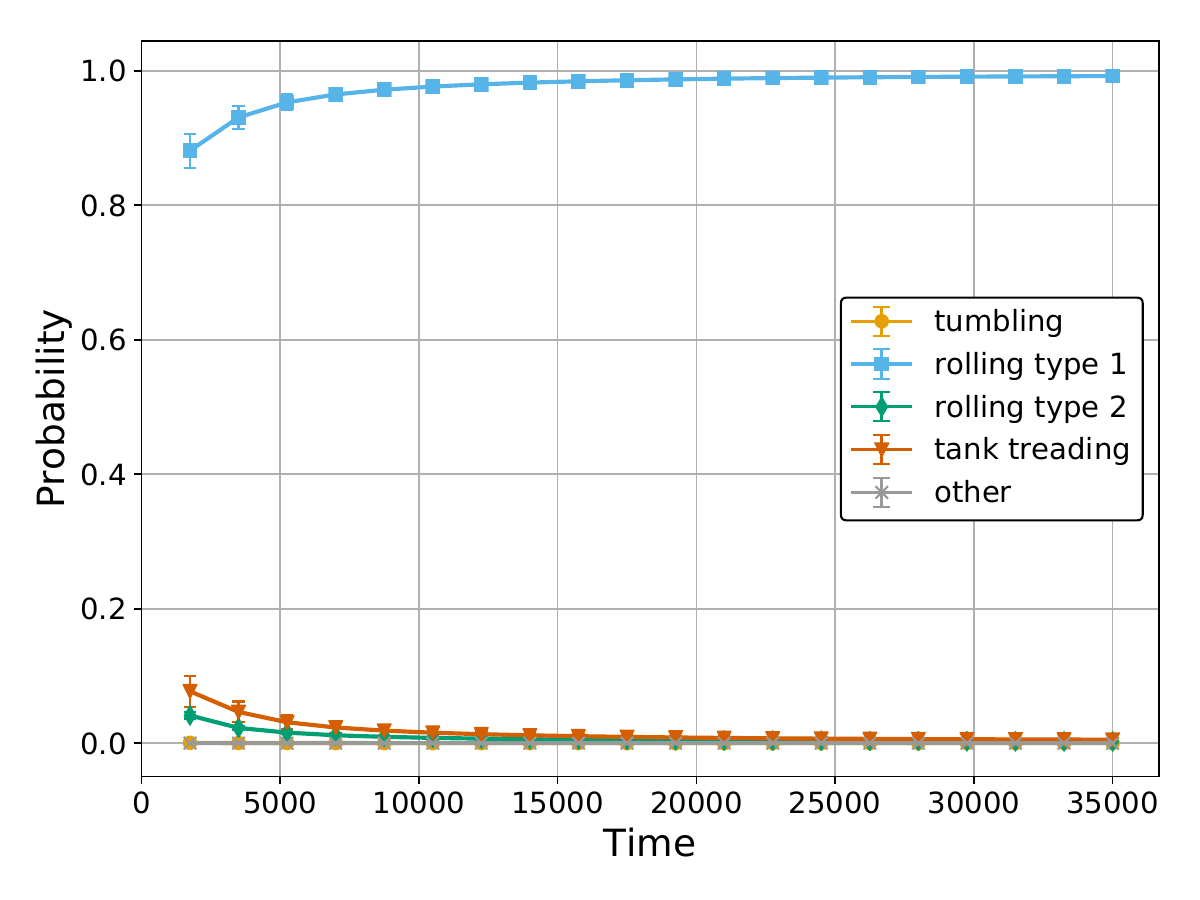}
        \put(2,77){\normalsize\sffamily (a)}
    \end{overpic}
\end{subfigure}
\hfill
\begin{subfigure}[t]{0.48\textwidth}
    \centering
    \begin{overpic}[width=\textwidth]{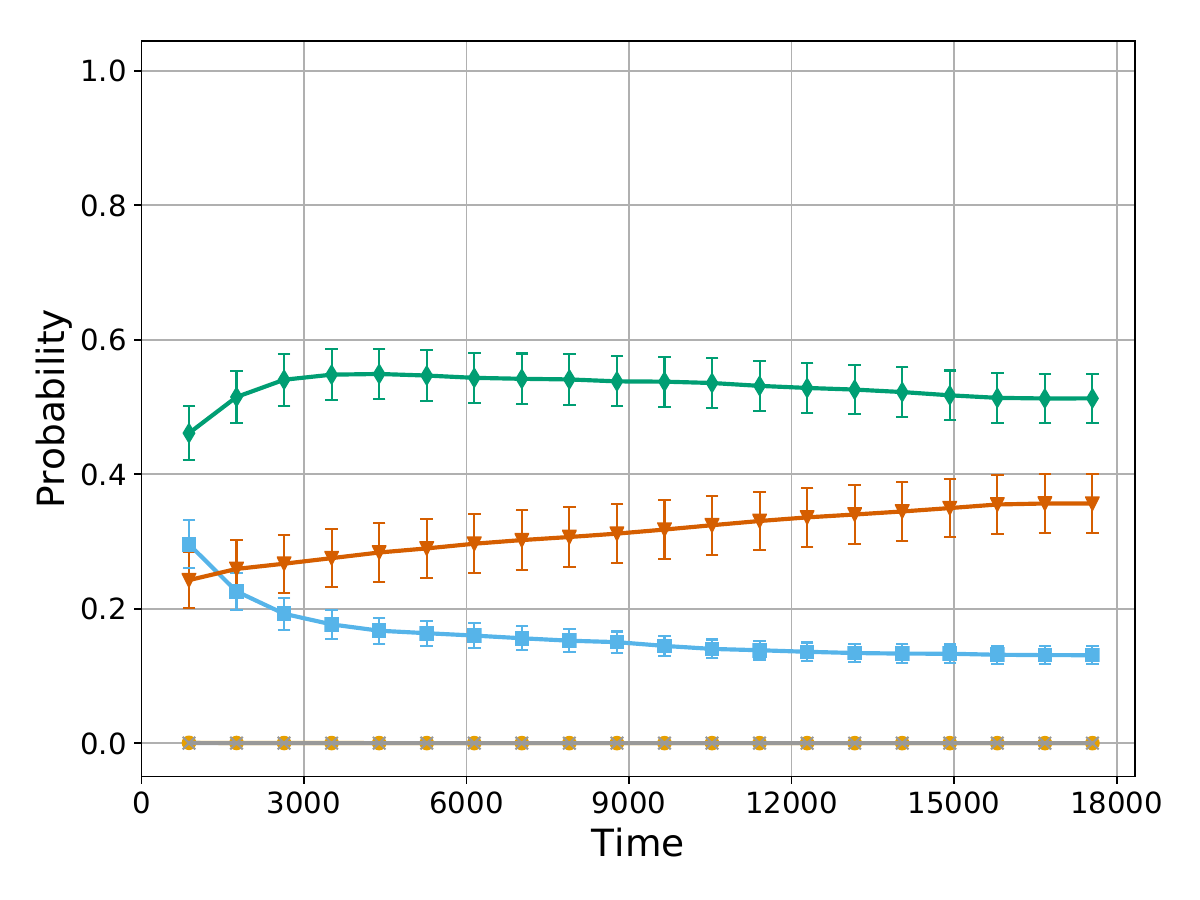}
        \put(2,77){\normalsize\sffamily (b)}
    \end{overpic}
\end{subfigure}

\vspace{0.5em}

\begin{subfigure}[t]{0.48\textwidth}
    \centering
    \begin{overpic}[width=\textwidth]{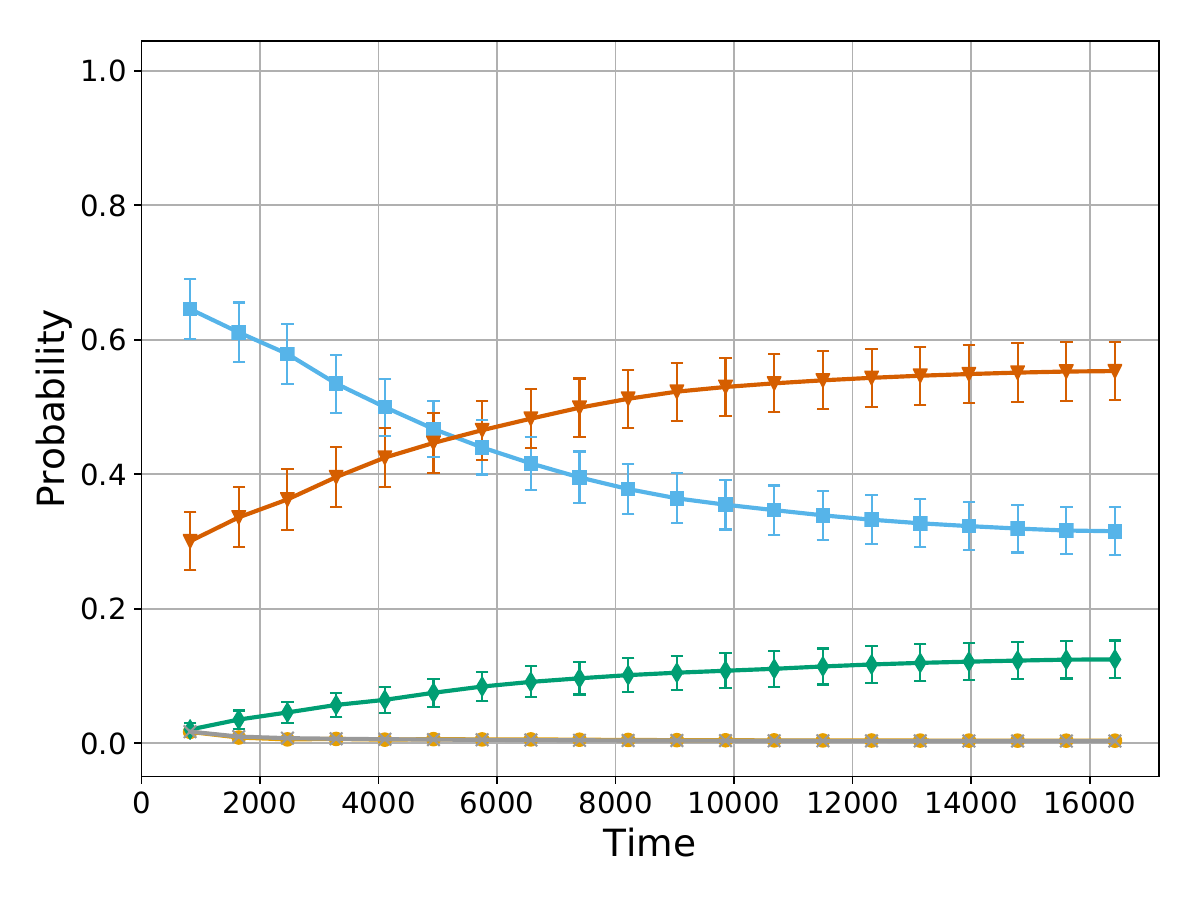}
        \put(2,77){\normalsize\sffamily (c)}
    \end{overpic}
\end{subfigure}
\hfill
\begin{subfigure}[t]{0.48\textwidth}
    \centering
    \begin{overpic}[width=\textwidth]{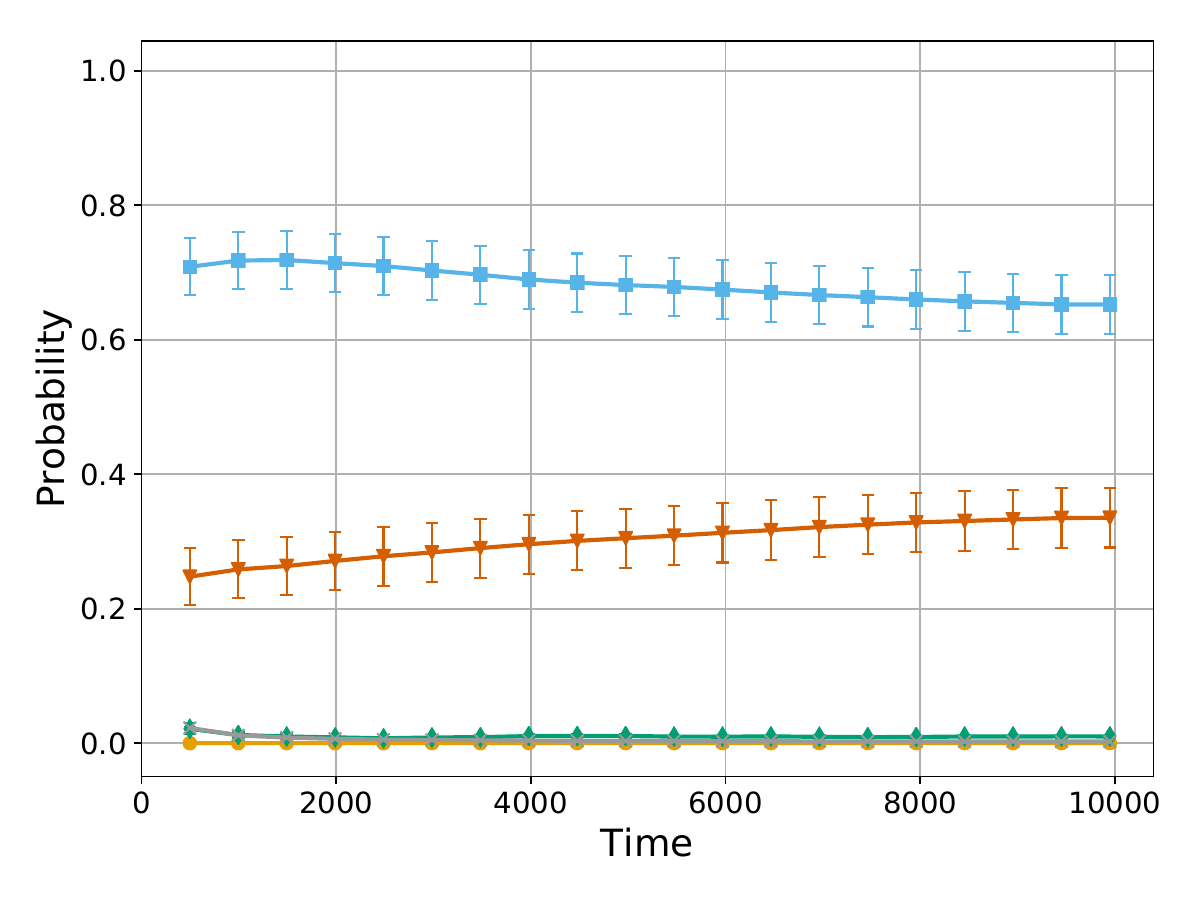}
        \put(2,77){\normalsize\sffamily (d)}
    \end{overpic}
\end{subfigure}

\vspace{0.5em}

\begin{subfigure}[t]{0.48\textwidth}
    \centering
    \begin{overpic}[width=\textwidth]{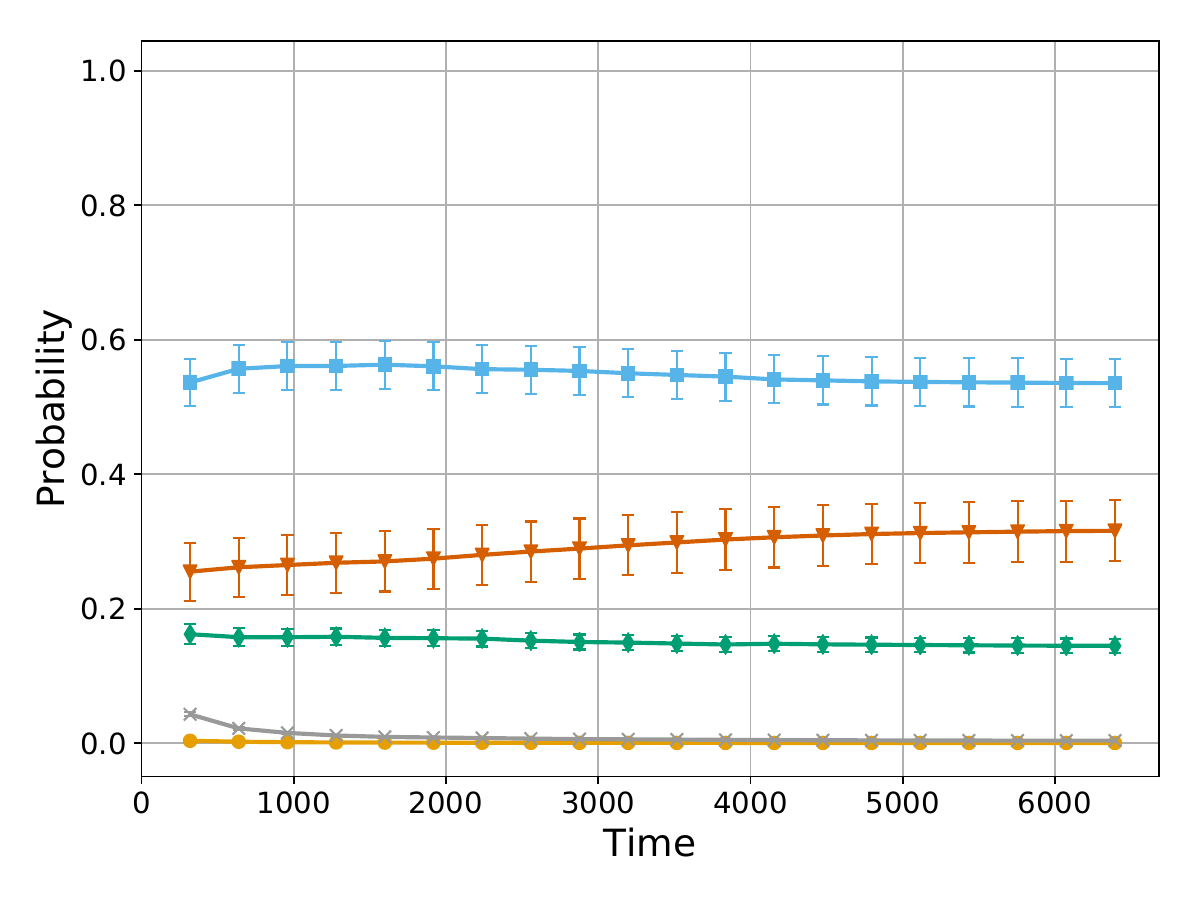}
        \put(2,77){\normalsize\sffamily (e)}
    \end{overpic}
\end{subfigure}
\hfill
\begin{subfigure}[t]{0.48\textwidth}
    \centering
    \begin{overpic}[width=\textwidth]{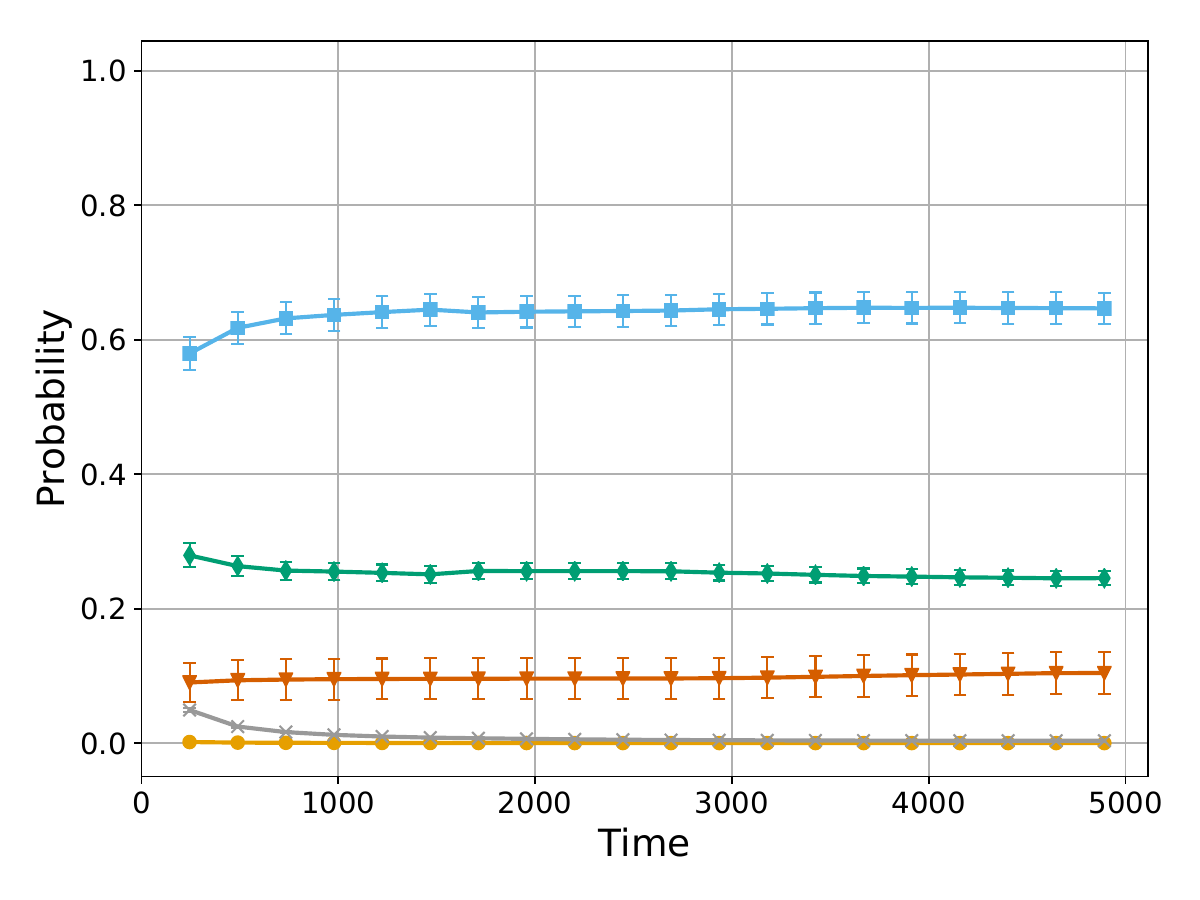}
        \put(2,77){\normalsize\sffamily (f)}
    \end{overpic}
\end{subfigure}

\caption{\justifying Cumulative probabilities of different dynamic behaviors over time for (a) \textup{[3]}catenane, (b) \textup{[5]}catenane, (c) \textup{[7]}catenane, (d) \textup{[9]}catenane, (e) \textup{[11]}catenane, and (f) \textup{[13]}catenane. Averages and uncertainies are determined from 100 independent simulations.}
\label{fig:modes_vs_time_odd}
\end{figure*}

\begin{center}
\begin{table}[h]
\centering
\caption{Probabilities of dynamic behaviors of even polycatenanes.}
\label{tab:even_catenantes}
\setlength{\tabcolsep}{4pt}
\begin{tabular}{ccccccccccc}
& \multicolumn{2}{c}{Tumbling} & \multicolumn{2}{c}{Rolling Type 1} & \multicolumn{2}{c}{Rolling Type 2} & \multicolumn{2}{c}{Tank Treading} & \multicolumn{2}{c}{Other} \\
Polymer & Simulation & MSM & Simulation & MSM & Simulation & MSM & Simulation & MSM & Simulation & MSM \\
\toprule
\textup{[2]}catenane & 0.0000 & 0.0000 & 0.9985 & 1.0000 & 0.0015 & 0.0000 & 0.0000 & 0.0000 & 0.0000 & 0.0000 \\
\midrule
\textup{[4]}catenane & 0.0000 & 0.0000 & 0.9904 & 0.0000 & 0.0096 & 1.0000 & 0.0000 & 0.0000 & 0.0000 & 0.0000 \\
\midrule
\textup{[6]}catenane & 0.0000 & 0.0000 & 0.7276 & 0.3080 & 0.2724 & 0.6920 & 0.0000 & 0.0000 & 0.0000 & 0.0000 \\
\midrule

\textup{[8]}catenane & 0.2528 & 0.2367 & 0.6216 & 0.6284 & 0.1248 & 0.1348 & 0.0000 & 0.0000 & 0.0008 & 0.0000 \\
\midrule

\textup{[10]}catenane & 0.0067 & 0.0008 & 0.8711 & 0.8632 & 0.1171 & 0.1360 & 0.0000 & 0.0000 & 0.0051 & 0.0000 \\
\midrule

\textup{[12]}catenane & 0.0000 & 0.0000 & 0.8789 & 0.8838 & 0.1136 & 0.1130 & 0.0060 & 0.0032 & 0.0015 & 0.0000 \\
\bottomrule

\end{tabular}
\end{table}
\end{center}

\begin{figure}[p]
    \centering    \includegraphics[width=0.75\textwidth,keepaspectratio]{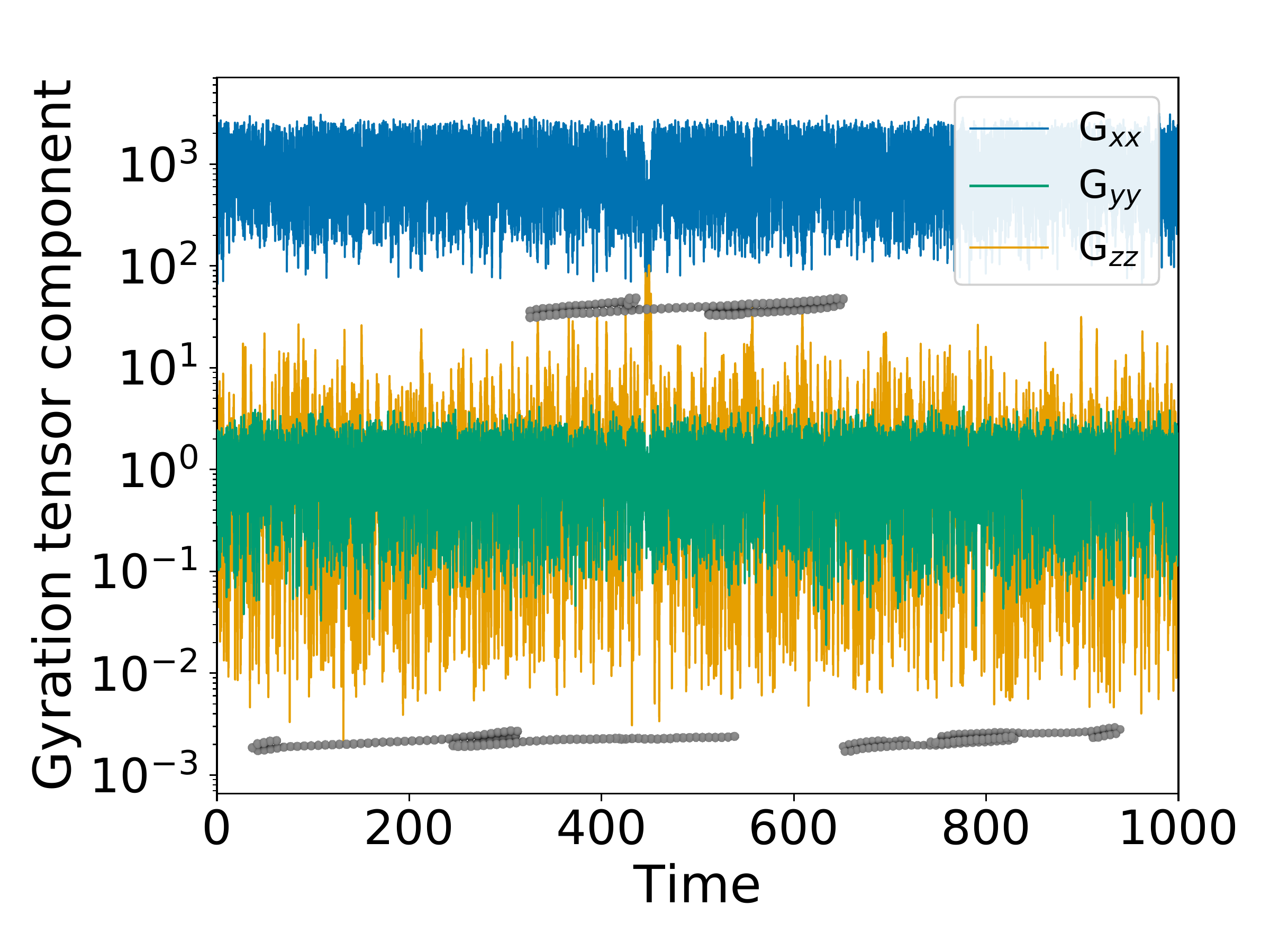}
        \caption{Gyration tensor components of a 100-bead linear polymer in strong shear flow.
        Large fluctuations in the components indicate stretch-tumble cycles and insets show typical configurations during these cycles.}
    \label{SI/gyration_linear}
\end{figure}

\begin{figure}[t]
 \includegraphics[width=\textwidth,keepaspectratio]{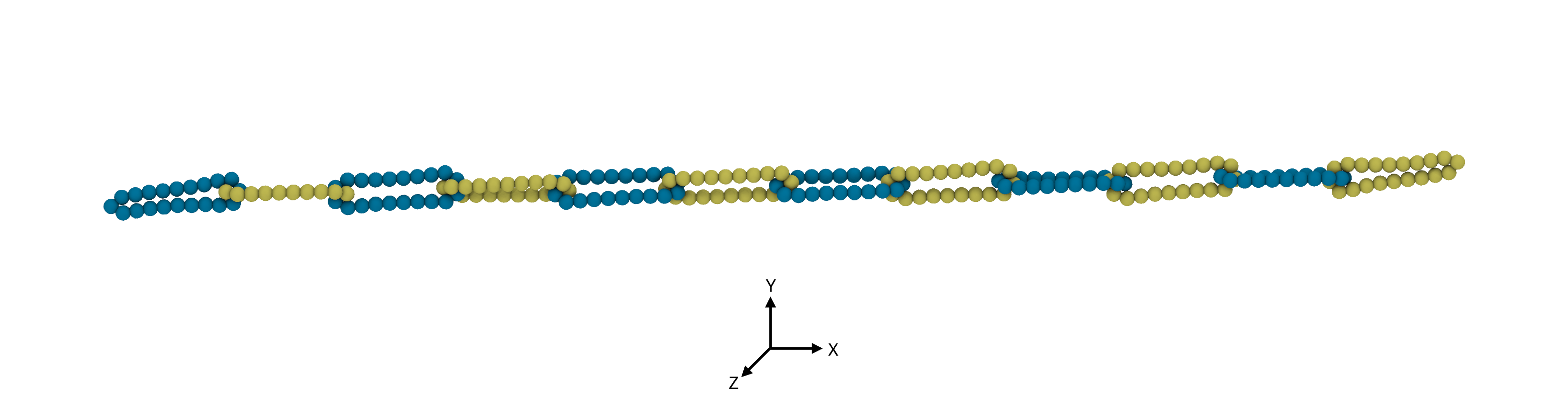}
        \caption{Trajectory snapshot of the [12]catenane while tank treading.  The outer 3 rings on each side are in a typical odd polycatenane tank treading configuration and serve to anchor and stretch the inner 6 rings, which twist out of the flow-gradient and flow-vorticity planes to accommodate the outer rings.}
    \label{SI/12_rings}
\end{figure}

\end{document}